\documentclass[11pt]{article}
\usepackage{times}
\usepackage{graphicx}
\usepackage{color}
\usepackage{amssymb}
\usepackage{amsmath}
\usepackage{ifthen}
\usepackage{calc}

\newtheorem{theorem}{Theorem}

\newcommand{\dist}[2]{\ensuremath{\mathsf{err\mbox{-}mean}\ifthenelse{\equal{#1}{}}{}{(#1,\,#2)}
}}
\newcommand{\wmean}[2]{\ensuremath{\mathsf{mid}_{#1}\ifthenelse{\equal{#2}{}}{}{(#2)}}}
\newcommand{\wmeandown}[2]{\ensuremath{\mathsf{mid}_{\downarrow #1}\ifthenelse{\equal{#2}{}}{}{(#2)}}}
\newcommand{\wmeanup}[2]{\ensuremath{\mathsf{mid}^{\uparrow #1}\ifthenelse{\equal{#2}{}}{}{(#2)}}}

\newcommand{\setmean}[1]{\ensuremath{\mathsf{mid}_\infty(#1)}}

\newcommand{\Strictdown}[2]{\ensuremath{\mathsf{strict}_{\downarrow{#1}}\ifthenelse{\equal{#2}{}}{}{(#2)}}}
\newcommand{\Strictup}[2]{\ensuremath{\mathsf{strict}^{\uparrow {#1}}\ifthenelse{\equal{#2}{}}{}{(#2)}}}

\newcommand{\lex}[2]{\ensuremath{\mathsf{lex}_{#1} \ifthenelse{\equal{#2}{}}{}{(#2)}}}

\newcommand{\nlog}[2]{\ensuremath{\Theta\left(n\ifthenelse{\equal{#1}{1}}{}{^{#1}}\ifthenelse{\equal{#2}{0}}{}{\log\ifthenelse{\equal{#2}{1}}{}{^{#2}} n}\right)}}

\newlength{\Ainlength}
\newlength{\Aintemp}
\newcommand{\Ain}[1]{\setlength{\Aintemp}{0.5in}\addtolength{\Aintemp}{#1\Ainlength} \hspace*{\Aintemp}}

\begin{document}

\begin{center}
{\Large \textbf{Best $L_p$ Isotonic Regressions,} $\mathbf{p \in \{0, 1, \infty\}}$}
\bigskip

{\large Quentin F. Stout}\\[\medskipamount]
Computer Science and Engineering, University of Michigan\\
qstout@umich.edu~~~~
www.eecs.umich.edu/{\,\raisebox{-0.5ex}{\textasciitilde}qstout/}

\end{center}
\bigskip

\noindent \textbf{Abstract:~}
Given a real-valued weighted function $f$ on a finite dag, the $L_p$ isotonic regression of $f$, $p \in [0,\infty]$, 
is unique except when $p \in [0,1] \cup \{\infty\}$.
We are interested in determining a ``best'' isotonic regression for $p \in \{0, 1, \infty\}$, where by best we mean a regression satisfying stronger properties than merely having minimal norm.
One approach is to use strict $L_p$ regression, which is the limit of the best $L_q$ approximation as $q$ approaches $p$, and another is lex regression, which is based on lexical ordering of regression errors.
For $L_\infty$ the strict and lex regressions are unique and the same.
For $L_1$, strict $q \scriptstyle\searrow 1$ is unique, but we show that $q \scriptstyle\nearrow 1$ may not be,
and even when it is unique the two limits may not be the same.
For $L_0$, in general neither of the strict and lex regressions are unique, nor do they always have the same set of optimal regressions, but by expanding the objectives of $L_p$ optimization to $p < 0$ we show $p$ {\scalebox{0.7}{$\nearrow$}} $0$ is the same as lex regression.
We also give algorithms for computing the best $L_p$ isotonic regression in certain situations.
\medskip 

\bigskip

\section{Introduction} \label{sec:Intro}

This paper considers isotonic regression (also known as monotonic regression) on an arbitrary finite dag $G=(V,E)$.
A real-valued function $h$ on $G$ is \textit{isotonic} if for all vertices $u, v \in V$, if $u \prec v$ (i.e., if there is a path from $u$ to $v$) then $h(u) \leq h(v)$, i.e., it is a weakly monotonic function from $G$ into $\Re$.
We are given a weighted function $(f,w)$ on $V$, where $f$ is a real-valued function and $w$ is the positive real-valued weights, and wish to produce a real-valued isotonic function $g$ that is closest to $f$.
If all the weights are the same then we say that the function is unweighted.
Isotonic regression is a form of nonparametric regression, and hence very useful when parametric assumptions are unwarranted.
Currently Google lists tens of thousands of results for a search on "isotonic regression".
They have become quite important in data science and machine learning, with all of the major software packages in these areas including algorithms to compute them.
An extensive overview of such applications appears in~\cite{Canoetal_MonoClass19}.

Normally the distance between $f$ and $g$ is measured in terms of the weighted $L_p$ distance, i.e., for $1 \leq p < \infty$ is
$$
(\sum_{v \in V}\, w(v) \!\cdot\! |f(v) - g(v)|^p)^{1/p} 
$$
Finding a $g$ that minimizes this is the same as finding one that minimizes the same sum without taking the $1/p$ root.
This then extends to $0 < p < 1$, where there is no longer a true norm, but there is an F-norm.
We also extend to $p=0$ (the Hamming distance) and $p=\infty$.
Thus  the goal is to minimize
$$
\begin{array}{ll}
\sum_{v \in V}\, w(v) \!\cdot\! \mathbf{1}(f(v) \neq g(v)) & p=0 \medskip \\
 \sum_{v \in V}\, w(v) \!\cdot\! |f(v) - g(v)|^p & 0 <  p  < \infty \medskip \\ 
  \max_{v \in V}\, w(v) \!\cdot\! |f(v) - g(v)|  & p=\infty\\
\end{array}
$$
among all isotonic functions $g$ on $G$.
If $g$ is an isotonic function minimizing this quantity then we say $g$ is an $L_p$ isotonic regression of $f$.

An important special case is when $G$ is a simple linear order and $f$ is a non-increasing function.
In this case the best isotonic regressions will be constant functions, and the value of any constant minimizing the regression error is a \textit{weighted $L_p$ mean}.
For weighted means the location of the function values is solely determined by the function values, so we often just consider the values of $f$ without specifically noting which vertex has which function value, i.e., it reduces to considering the weighted mean of a multiset of weighted values.
Let \wmean{p}{S} denote the set of weighted $L_p$ means of $S$.

For $p \in (1,\infty)$ it is well-known that an $L_p$ regression $g$ is always unique due to the strict convexity of the objective function, and similarly \wmean{p}{S} is unique for any $S$.
For $p \in [0,1]$  the objective function is not strictly convex and $L_p$ regression, nor \wmean{p}{S}, need not be unique.
E.g., for all such $p$, for unweighted data 1, 0 on a linear order both 1, 1 and 0, 0 are optimal $L_p$ isotonic regressions, and are the elements in \wmean{p}{} for $p \in (0,1)$.
For $p \in (0,1)$ the objective is in fact strictly concave.
For $p=1$ the objective behavior is a bit more subtle, and \wmean{1}{} is either a single point or an interval.
For example, $\wmean{1}{\{0,1\}}= [0,1]$.
For $p=0$ \wmean{0}{S} is the set of values with maximum total weight.

This paper is concerned with $p =0, 1, \infty$, values for which the isotonic regression need not be unique.
A natural question is whether there are ``best'' $L_0$, $L_1$ and $L_\infty$ isotonic regressions.
Section~\ref{sec:definitions} gives two approaches to this question.
One is via limits of $L_p$ regressions as $p \rightarrow $ 0, 1, or $\infty$; the other is via lexical ordering of the regression values, useful for $L_0$ and $L_\infty$.
Section~\ref{sec:Linfty} discusses the fact that for $L_\infty$ the limit from below and a lexical approach give the same regression,
Section~\ref{sec:L1} shows that for $L_1$ the two limit approaches give different results, and
Section~\ref{sec:L0} shows that for $L_0$ the limit from above and a lexical approach give different results, but if $L_p$ optimization is extended to $p<0$ then the limit from below and the lexical approach are the same.

Section~\ref{sec:linear} gives algorithms for computing these regressions on linear orders using partitioning and ``pool adjacent violators'' (PAV), and Section~\ref{sec:Conclude} gives some concluding remarks.

\section{Definitions} \label{sec:definitions}

To determine best isotonic regressions, one approach is by looking at the limiting behavior of $L_q$ regression as $q$ approaches $p$.
These limits are sometimes called \textit{strict} isotonic regressions.
Given a weighted function $f$ on a dag, for $L_\infty$ we consider the limit of $L_p$ regressions as 
$p$ \scalebox{0.8}{$\nearrow$} $\infty$, denoted \Strictup{\infty}{f} (Section \ref{sec:Linfty}); 
for $L_1$ we consider \Strictup{1}{f} and \Strictdown{1}{f} (Section \ref{sec:L1}); and
for $L_0$ consider \Strictdown{0}{f} (Section \ref{sec:L0}).
Given the strict convexity of the objective functions being optimized,  for  $1 < p <\infty$ it is not difficult to show that \Strictup{p}{f} and \Strictdown{p}{f} are the unique $L_p$ isotonic regression.
Many problems have been analyzed using \Strictdown{1}{f} and  \Strictup{\infty}{f}~\cite{DLG_LinfPolya83,EggerTaylorPolya1ConvRate93,HLMT_L1Polya88,QFOMMB_PolyaL1}.
The use of \Strictup{\infty}{f} is sometimes called the \textit{Polya approach}, and, less frequently, the use of
\Strictdown{1}{f} is called the \textit{Polya-1 approach}.
While here \Strictup{\infty}{f} and \Strictdown{1}{f} are unique, for some classes of functions on infinite sets this is not true~\cite{BadL1}.

We also consider another approach that has been used to determine ``best''' $L_\infty$ and $L_0$ regressions.
For an isotonic regression $g$ of an unweighted function $f$ on a DAG of $n$ vertices, take the regression errors of $g$ at the vertices and sort them in decreasing order, giving a vector of $n$ entries.
Order all such vectors lexically, and let $v$ be the minimal vector in this ordering.
While there are infinitely many isotonic regressions of $f$ it can be shown that $v$ is well-defined and
corresponds to a unique isotonic regression~\cite{QStrictLinfty} $h$ which minimizes the $L_\infty$ error.
$h$ has been called the minimizing lex regression or similar terms.
Here we denote it as \lex{\infty}{f}.
For $L_0$ on unweighted functions use the same process, but now have the vector entries in increasing vertex error.
Once again there is a well-defined minimal vector $v$ in the lexical ordering of the vectors, though
there may be multiple isotonic regressions corresponding to $v$ (Sec.\ \ref{sec:L0}), each of which is an $L_0$ optimal isotonic regression of $f$.
We denote these regressions as \lex{0}{f}, though they have also been called \textit{strong $L_0$ regression}~\cite{QL0Secondary}.

For weighted values the definition of the lexical orderings have to be generalized a bit.
For \lex{\infty}{f,w}, for isotonic regressions $g, h$ define $g \prec h$ iff there is an $\alpha > 0$ such that
$$\sum \{w(v): |g(v)\!-\!f(v)| \geq \alpha,~ v \in V\}  < \sum \{w(v): |h(v)\!-\!f(v)| \geq \alpha,~ v \in V\}$$
and for all $\beta > \alpha$,
$$\sum \{w(v): |g(v)\!-\!f(v)| \geq \beta,~ v \in V\}  = \sum \{w(v): |h(v)\!-\!f(v)| \geq \beta, ~v \in V\}$$

\noindent
For \lex{0}{f,w}, for isotonic regressions $g, h$ define $g \prec h$ iff there is an $\alpha > 0$ such that 
$$\sum \{w(v): |g(v)\!-\!f(v)| \leq \alpha, ~v \in V\}  > \sum \{w(v): |h(v)\!-\!f(v)| \leq \alpha, ~v \in V\}$$
and for all $\beta < \alpha$,
$$\sum \{w(v): |g(v)\!-\!f(v)| \leq \beta, v \in V\}  = \sum \{w(v): |h(v)\!-\!f(v)| \leq \beta, ~v \in V\}$$
As before, \lex{\infty}{f,w} and \lex{0}{f,w} are initial elements in their orderings.

It had previously been shown that $\Strictup{\infty}{f}=\lex{\infty}{f}$~\cite{QStrictLinfty},
while in Section~\ref{sec:L0} we show that \Strictdown{0}{f} is not always the same as \lex{0}{f}.
In Section~\ref{sec:L1} we show how to approximate \Strictdown{1}{} and that in general it is different than \Strictup{1}{}.
In the Appendix we show that if an $L_\infty$ isotonic regression algorithm satisfies monotonicity and level set trimming then it is in fact \lex{\infty}{}.
A slightly improved algorithm for finding the \lex{\infty}{} regression is given in Section~\ref{sec:Linfty}.

For $p \in [0,\infty)$ let \wmeandown{p}{S} be $\lim_{q \searrow p} \wmean{q}{S}$, and for $p \in (0,\infty]$, \wmeanup{p}{S} denotes $\lim_{q \nearrow p} \wmean{q}{S}$.

If $g$ is an isotonic function on $G$ then a set $S \subseteq V$ is a \textit{level set} if it is a maximal weakly connected set where all the values of $g$ are the same.
It is straightforward to show that for any $p \in (0,\infty)$, if $g$ is an optimal $L_p$ isotonic regression then for any level set $S$ of $g$, $g$'s value on the level set is in \wmean{p}{S}.

Given a function $f$ on a dag $G=(V,E)$, say that $u,v \in V$ are a \textit{violating pair} if $u$ precedes $v$ in $G$ but $f(u) > f(v)$, i.e., they violate the isotonic condition.
The fastest known algorithms for determining \lex{\infty}{} and \lex{0}{} on a general dag are based on first finding all violating pairs.
This can be done in time linear in the time to find the transitive closure of $G$, which can be done in 
$O(nm, n^\omega)$ time, where $n = |V|$, $m=|E|$, and $\omega$ is such that binary matrix multiplication can be done in $O(n^\omega)$ time.
However, there are some orderings where the violating pairs can be found more quickly.
For example, suppose the dag has vertices where the ordering relationship can be determined directly from pairwise comparisons of the vertices' labels and the edges are not explicitly given.
The violating edges can therefore be determined in $O(n^2)$ time.
An example of this is where vertrex labels are strings and $u \prec v$ iff $u$ is a substring of $v$ (though a timing analysis should take the time of the comparisons into account since strings can be of varying size).
Another example is planar rectangles of arbitrary orientation and size, where $p \prec q$ iff $p$ is contained in $q$ (and, more generally, polyhedral containment in $d$-dimensional space).
A particularly important class where edges are given implicitly are points in $d$-dimensional space, where $u = (u_1,\ldots,u_d) \prec v=(v_1,\ldots,v_d)$ iff $u_i \preceq v_i$ for all $1 \leq i \leq d$.
This is sometimes known as \textit{domination ordering}, where $v$ dominates $u$, and it is also known as \textit{multidimensional ordering} or \textit{coordinate-wise} ordering.
For these orderings the transitive closure can be found in $\Theta(C + n \log^d n)$ time, where $C$ is the size of the transitive closure and where the implied constants depend on $d$.
Use of these facts to quickly find $L_0$ regressions appears in~\cite{QL0Secondary}.

\section{$L_\infty$ } \label{sec:Linfty}

The best $L_\infty$ regression has been defined both as \Strictup{\infty}{} (the Polya approach) and as \lex{\infty}{}.
A proof that these are the same, and the regression is unique, appears in~\cite{QStrictLinfty}.
To date apparently all efficient algorithms for determining the regression on finite dags are based on the \lex{\infty}{} definition.
For an arbitrary dag of $n$ vertices and $m$ edges a $O(\min\{nm,n^\omega\} + n^2 \log n)$ time algorithm appears in~\cite{QStrictLinfty} (where $\omega$ is such that boolean matrix multiplication can be performed in $O(n^\omega)$ time), and
the CompLexMin algorithm in~\cite{Yale_LipschitzLearn} computes it in $\Theta(nm)$ expected time.
Algorithm A below is a simplification of~\cite{QStrictLinfty} which should be the fastest in practice.

Algorithm A is essentially as follows: for each violating pair of vertices $(u,v)$ determine the weighted average $x = (w_uf(u) + w_v f(v))/(w_u+w_v)$, i.e., their $\mathsf{wmean}_p$ for any $1 \leq p \leq \infty$, and let their
{\sffamily \small mean\_err} be $w_u \!\cdot\! |f(u)-x|$, which is the same as $w_v \!\cdot\! |f(v)-x|$. 
Create a record which includes $u$, $v$, and the {\sffamily \small mean\_err} of their weighted function values.
Sort these records by the {\sffamily \small mean\_err} value in decreasing order, and then process them in sequence.
If the next record contains vertices $u \prec v$ and neither has had their regression value determined yet set the regression values to $x = \mathsf{wmean}$, set the upper bound of all predecessors of $v$ to the maximum of their previous value and $x$, and set the lower bound of all successors of $u$ to the minimum of their previous value and $x$.
If both have had their regression value determined already discard the record, and if one has had the regression value defined while the other hasn't the steps are shown in Algorithm A.
A proof of correctness is essentially the same as the proof of the algorithm in~\cite{QStrictLinfty}.

The total time of the algorithm is the time to find all violating pairs, the time to sort the records, and the time to process them.
For any violating pair $p \prec q$ the conceptual edge from $p$ to $q$ is used at most twice, once when $p$'s regression value is determined and once when $q$'s is.
Since the time to process the records is linear in the number of records, the sorting time dominates the remaining steps other than the time to find the violating pairs.
In terms of the original dag this is at most $\Theta(n^2 \log n)$, and thus A's worst-case time is $O(\min\{nm,n^\omega\} + n^2 \log n)$, i.e.,  the same as the time of the algorithm in~\cite{QStrictLinfty} though some of the steps are slightly faster.
Note that for dags such as those discussed at the end of Section~\ref{sec:definitions} the time for finding the violating pairs is $\Theta(n^2)$, in which case Algorithm A takes $\Theta(n^2 \log n)$ time.
Further, in terms of the number of violating pairs, $m^*$, the time to sort and process is $\Theta(m^*\log m^*)$ and thus the more in-order the data is the faster the later parts of the algorithm are.

\begin{figure*}
{\sffamily \small
\Ain{0} input: weighted data (f,w), lists of successors and predecessors for each vertex in dag G\\
\Ain{0} output: S = \lex{\infty}{\mathsf{G,f,w}}\\
\Ain{0} violators: array of (mean\_error,u,v)  for violating pairs u $\prec$ v, f(u) $>$ f(v)\\
\Ain{0} lowbd(v), upbd(v): lower and upper bounds on S(v)\\

\Ain{0} numviolate=0\\
\Ain{0} for every vertex v\\
\Ain{1}   lowbd(v) $=-\infty$;~  upbd(v) $=+\infty$;~ S(v) $=$ undefined\\
\Ain{1}   for every successor s of v\\
\Ain{2}      if f(v) $>$ f(s) then violators(numviolate)= (mean\_err(v,s), v, s);~ numviolate++\\

\Ain{0} sort violators by decreasing order of mean\_err. For ties sort by decreasing order of weight\\

\Ain{0} for i=0 to numviolate-1\\
\Ain{1}   (mean\_err,pred,suc)=violators(i)\\
\Ain{1}   if (S(pred) defined) $\vee$ (S(suc) defined) then cycle\\
\Ain{1}   wmean $=$ weighted mean of pred and succ\\
\Ain{1}   if wmean $\geq$ upbd(pred) then~ \{f(pred) is $\geq$ upbd(pred), no later mean is $<$ upbd(pred)\}\\
\Ain{2}     S(pred) $=$ upbd(pred)\\
\Ain{1}   if wmean $\leq$ lowbd(suc) then~ \{f(suc) is $\leq$ lowbd(suc), no later mean is $>$ lowbd(suc)\}\\
\Ain{2}     S(suc) $=$ lowbd(suc)\\
\Ain{1}  if (S(pred) undefined) $\wedge$ (S(suc) undefined) then~ \{low(suc) $\leq$ wmean $\leq$ high(pred)\}\\
\Ain{2}    S(pred) $=$ S(suc) $=$ wmean\\

\Ain{1} if S(pred) defined then\\
\Ain{2}   for every successor s of pred\\
\Ain{3}      lowbd(s) $=$ max\{lowbd(s),S(pred)\}\\
\Ain{1} if S(suc) defined then\\
\Ain{2}    for every predecessor p of suc\\
\Ain{3}       upbd(p) $=$ min\{upbd(p),S(suc)\}\\
\Ain{0} end for i\\

\Ain{0} for every vertex v\\
\Ain{1}    if S(v) undefined then\\ 
\Ain{2}       if f(v) $\geq$ upbd(v) then S(v)=upbd(v)\\
\Ain{2}       else if f(v) $\leq$ lowbd(v) then S(v)=lowbd(v)\\
\Ain{2}       else S(v)=f(v)
}
\bigskip \bigskip

\centerline{Algorithm A: Computing S=\lex{\infty}{G,f,w} ($=$ \Strictup{\infty}{G,f,w}) via Transitive Closure}

\end{figure*}

\section{$L_1$} \label{sec:L1}

Most published $L_1$ isotonic regression algorithms use a median data value as the value of a level set~\cite{AHKS_L1,Yale_AllLp,Pardalosetal94,RobertsonWright,Rote,ShiMinL195,QPartition,QLpviaL0},
greatly simplifying the search for optimal regression values.
Other regression values can also optimize the $L_1$ regression error, and have additional desirable properties.
However, computing them can be more complicated.
We will show that for a set $S$ of weighted real numbers, in general $\wmeandown{1}{S} \neq \wmeanup{1}{S}$, and hence in general $\Strictdown{1}{f} \neq \Strictup{1}{f}$.
When $f$ is unweighted data on a linear order, with values 1, 0, algorithms based on data values would result in either 0, 0 or 1, 1 as the isotonic regression.
Meanwhile, \Strictdown{1}{f} is 0.5, 0.5 and \wmeandown{1}{\{0,1\}} is \{0.5\}, while \Strictup{1}{f} is 0, 0 or 1, 1 and \wmeanup{1}{\{0,1\}} is \{0,1\}.
While some attention had been paid to \wmeandown{1}{}, apparently none had been to \wmeanup{1}{}.

One could also define versions of best $L_1$ isotonic regressions in terms of lexical properties.
Take the set of $L_1$ optimal regressions, order their regression errors as for \lex{\infty}{}, and take a regression corresponding to the first element in this ordering.
Denote this by \lex{1,\infty}{}.
For a set $S$ of weighted real numbers let \wmean{1,\infty}{S} be the regression of smallest $L_\infty$ error among \wmean{1}{S}.
For unweighted data $S = \{1, 1, 3, 7\}$: \wmean{1,\infty}{S} is 3; \wmean{\infty}{S} is 4;  \wmeandown{1}{S} is a value in (1,3) that is closer to 3 than to 1 (see Sec.~\ref{sec:strictdown1}); and \wmeanup{1}{S}$=1$ (see Sec.~\ref{sec:strictup1}).
A similar technique can be used for \lex{1,0}{}, though this is not always unique.
For example, for $S= \{1, 2, 3, 4\}$, \wmean{1,0}{S} = \{2, 3\} while \wmean{1}{S}$=[2,3]$.
While there is likely some interest in \lex{1,\infty}{} and \lex{1,0}{}, they won't be pursued any further here.

\subsection{$\mathsf{strict}_{\downarrow 1}$}   \label{sec:strictdown1}

Jackson~\cite{Jackson_Median} was apparently the first to determine \wmeandown{1}{S}.
He only considered unweighted data, but the extension to weighted data is straightforward.
If $S$ has a unique median value then that is  \wmeandown{1}{S}.
Otherwise, a nonempty $S$ must have unique values $a < b$ where both are weighted medians.
In this case \wmeandown{1}{S} is the unique value $c \in (a,b)$ such that 
$$\prod_{y_i \leq a} (c-y_i)^{w_i} = \prod_{y_i \geq b} (y_i - c)^{w_i}$$
or, equivalently,
$$\sum_{y_i \leq a}w_i \ln (c-y_i) = \sum_{y_i \geq b} w_i \ln(y_i-c)$$

Finding an $L_1$ isotonic regression where all level sets have data values and then using Jackson's formula on the regression values to determine the regression value of the level set does not always produce \Strictdown{1}{}.
For example, on a linear order with unweighted data values 0, -2, 2, 0 one optimal $L_1$ isotonic regression is 0, 0, 0, 0.
Applying Jackson's formula to this level set regression values would of course give the same values, 
but \Strictdown{1}{} is -1, -1, 1, 1.
Note that the level sets are not the same, let alone the regression values.
Jackson's formula needs to be applied to the original data values, not the regression values.

It appears that no previous algorithm has been published which determines \Strictdown{1}{} for isotonic regression on finite sets, though \Strictdown{1}{} isotonic regression has been examined in settings where the functions are integrable~\cite{HLMT_L1Polya88}, and has been examined in more general settings where the goal is to find a closest point on a closed convex set~\cite{LandersRogge_L1approx}.
To find a close approximation of \Strictdown{1}{} one can determine a $p > 1$ close enough to 1 so that the $L_p$ isotonic regression is sufficiently close to \Strictdown{1}{}.
Then approximate the $L_p$ isotonic regression.
To simplify, assume we want an approximation to within $2\delta$ pointwise, $\delta > 0$, and each of these two steps has its parameters chosen to produce an approximation within $\delta$ at each vertex.

Suppose all the function values are integers in the range $[-h,h]$ and the weights are integers in the range $[0,W]$.
If the values and weights aren't integers then just round and do pointwise approximation to within $\delta/2$.
For a given $p>1$, for any set $S$ of $n$ or fewer weighted values, the largest difference between the weighted $L_1$ mean  of $S$ and the $L_p$ mean occurs if all vertices have weight $W$ and half have value $h$ while the other half have value $-h$.
The difference in means is $nWh - (nWh^p)^{1/p}$, which is $\leq \delta$ iff $nW-(nW)^{1/p} \leq \delta/h$.
Rearranging and taking the log of both sides, this will be true if $1/p \geq \ln(nW-\delta/h)/\ln(nW)$, which will hold if $p$ is sufficiently close to 1.
Note that this approach can be used for arbitrary $L_1$ regression, not just isotonic regression~\cite{LandersRogge_L1approx}.

To find an approximation to the $L_p$ isotonic regression there are several approaches.
In~\cite{Yale_AllLp} they use interior point methods, while~\cite{QPartition,QLpviaL0} iteratively use an algorithm for weighted $L_1$ binary-valued isotonic regression to converge to an approximation in a logarithmic number of steps.
For a connected dag with $n$ vertices and $m$ edges, the algorithm in~\cite{Yale_AllLp} produces an approximation within $\delta$ in $O(m^{1.5} \log^2 n \log (n/\delta))$ time, while the approach in~\cite{QPartition} depends on the dag.
If $K = nhW$ then the time is $\Theta(n \log K)$ if the dag is linear, a tree, or a 2-dimensional grid; $\Theta(n \log n \log K)$ if the dag is arbitrary points in 2-dimensional space with domination ordering; $\Theta(n^2 \log n \log K)$ for a $d$-dimensional grid, $d \geq 3$, and $\Theta(n^2 \log^d n \log K)$ for arbitrary points in $d$-dimensional space with domination ordering (for these results the implied constants depend upon $d$).

The fastest known algorithms for arbitrary dags are approximations based on algorithms for $L_0$~\cite{QLpviaL0}, which are in turn based on flow algorithms~\cite{RadeBM_MaxIS}.
Given a flow algorithm $F$, let $\mathcal{F}(n,m,nU)$ denote the time to solve an integer-valued flow on a graph with $n$ vertices, $m$ edges, and integer edge flow limits in $[nU]$ for some positive integer $U$.
Let $G' = (V',E')$ be a violator dag of a dag $G = (V,E)$, where $G'$ has $n'$ vertices and $m'$ edges.
Then for $1 < p < \infty$, given a weighted function $(f,w)$ on $G$ with integer weights in $[0,U]$, and given $G'$, for $\delta > 0$ an isotonic function within $\delta$ of the $L_p$ isotonic regression of $f$ can be found in $O(\mathcal{F}(n',m', nU) \log K)$ time, where $K = \left(\max_{v \in V} f(v)  - \min_{v \in V} f(v)\right)/\delta$ and where the implied constants in the O-notation depend on $p$.
A recent improvement in flow algorithms~\cite{Flow_That} reduces their time to  $\tilde{O}(m)$ if $K$ grows at most polynomially in $n$, and thus the total time is bounded by this plus the time to construct the violator graph.

\subsection{$\mathsf{strict}^{\uparrow 1}$}   \label{sec:strictup1}

Minimizing $ \sum_{v \in V}\, w(v) \!\cdot\! |f(v) - g(v)|^p $ when $p \in (0,1)$ is different than minimizing it when $p \geq 1$ because for $p < 1$, $|f(v)-y|^p$ is a concave function of $y$ on $y \leq f(v)$ and on $y \geq f(v)$.
This in contrast to it being convex throughout the full range of $y$ when $p \geq 1$, and strictly convex when $p > 1$.
Given the set $S = \{f(v): v \in V\}$, let $a, b$ two elements of $S$ where $a < b$ and there are no elements of $S$ in $(a,b)$.
Then for $0 < p < 1$, the $L_p$ error of using $y$ as the \wmeanup{1}{S} is a concave function on $[a,b]$.
This because each term of the error sum is concave in $y$ and the sum of concave functions is concave.
Thus the minimum on $[a,b]$ is achieved when $y = a$ or $y=b$.
Since this is true for any consecutive pair of data values, the minimum, i.e.\ \wmeanup{1}{S}, is achieved at one of the data values.

First we show that the value is a median.
If the regression value is chosen to be $\hat{y}$, then, letting $p=1-\epsilon$, for $\epsilon > 0$ sufficiently close to 0, the Taylor expansion series shows that the regression error is
$$
\sum_{v \in V} w_v|f(v)\!-\!\hat{y}|\, (1 + \mathrm{~l.o.t.})
$$
The sum with the final factor in each term being 1 instead of (1 + l.o.t.) is the sum under the $L_1$ norm, so it is the medians which minimize it.
Since the minimizer for \wmeanup{1}{S} must be a data value, it must be a median of $S$, either the unique median if $S$ has a unique median, and otherwise there are unique values $a < b$ in $S$ which are medians of $S$. Thus \wmeanup{1}{S} is either $a$ or $b$.
To determine if it is $a$, let $c(v)=|f(v)-a|$ and $d(v)=|f(v)-b|$.
The relevant question is if 
$$
\sum_{v \in V,\, v \neq a} w(v)c(v)^{1-\epsilon} ~<~ \sum_{v \in V,\, v \neq b} w(v)d(v)^{1-\epsilon}
$$
is true for $\epsilon$ sufficiently small.
Using Taylor expansions, and the fact that $\sum_{v \in V} w(v) d(v) = \sum_{v \in V} w(v) e(v)$ (since both are medians), this will be true if
$$
\sum_{v \in V,\, v \neq a} w(v) c(v) \ln c(v) ~< \sum_{v \in V,\, v \neq b}  w(v) d(v) \ln d(v) \hspace{0.15in}
$$
or, equivalently,
$$
\prod_{v \in V,\, v \neq a} c(v)^{w(v)c(v)}  ~<  \prod_{v \in V,\, v \neq b} d(v)^{w(v)d(v)}
$$
\wmeanup{1}{S} will be $b$ if the inequality is reversed.
If these are equal then one can go to the next order term in the Taylor expansion, etc.

Note that if there is not a unique median value then  \wmeandown{1}{S} cannot be one of the data values, while \wmeanup{1}{S} is always a data value.
Thus in general \Strictup{1}{} is not the same as \Strictdown{1}{}.
Further, they need not have the same level sets.
On a linear order, if the unweighted data values are 1, 1, -10, -11, 0, 0, -2, -3, then for the first 4 entries \wmeanup{1}{} is 1, 1, 1, 1 and for the second 4 is 0, 0, 0, 0, while \wmeandown{1}{} is negative on the first 4 and negative but somewhat larger on the second 4.
For \wmeanup{1}{} the 8 elements have to be merged into a single level set with value in \{-2,0\}, while for \wmeandown{1}{} the level sets are not merged.

There does not appear to be any published algorithm for determining isotonic regressions when $0<p<1$, so currently one cannot approximate \Strictup{1}{} by using a published $L_p$ approximation algorithm for $p<1$ sufficiently close to 1.

\section{$L_0$} \label{sec:L0}

The definition of the best $L_0$ regression as being \lex{0}{} apparently first appeared in~\cite{QL0Secondary}, where in the first version of that paper it was called strong $L_0$ isotonic regression.
In general \lex{0}{} is the not same as \Strictdown{0}{} since for 11, 8, 5, 2, 0, \lex{0}{} is 2 and \Strictdown{0}{} is 5.
However, we can consider \Strictup{0}{} by using an extension to negative $p$. 
For $0 > p > -\infty$  let 
$$
|| f - g ||_p = (\sum_{v \in V}\, w(v) \!\cdot\! |f(v) - g(v)|^p)^{1/p}
$$
\noindent This is not a norm, but we have the same goal as before, namely in minimizing it.
Since $p < 0$, this requires \textit{maximizing}~ $\sum_{v \in V}\, w(v) \!\cdot\! |f(v) - g(v)|^p$.
Viewing the set of optimal $g$ as a function of $p$, we define \Strictup{0}{} as the set
$\lim_{p{\scalebox{0.7}{$\nearrow$}} 0}$.
We will show that this is \lex{0}{}.

 To explain our approach assume all the weights are 1. 
 The approach works for arbitrary positive weights, but unit weights simplifies the explanation.
 Recall that to determine \lex{0}{} for each $L_0$ regression we consider its errors at the vertices in nondecreasing order.
 For regression $A$ denoted this list by $a_1, \ldots, a_n$, where $a_1 \leq  a_2 \ldots \leq a_n$,
 and for regression $B$ denote its list by $b_1, \ldots, b_n$, where $b_1 \leq b_2 \ldots  \leq b_n$.
 We will show that if $A$'s list precedes $B$'s in lexical order then $A$ has smaller $L_p$ error for $p$ sufficiently close to 0, $p < 0$.
 
 $A$'s precedes $B$'s iff there is an index $i$, $0 \leq i \leq n-1$, such that $a_j = b_j$ for $1\leq j \leq i$, and $a_{i+1} < b_{i+1}$. 
 We can ignore the values of $a_j$ and $b_j$ for $j \leq i$ since they contribute the same to each sum.
 Let $A_p = \sum_{k=i+1}^n a_k^{1/p}$ and $B_p=\sum_{k=i+1}^n b_k^{1/p}$.
 Then the sum over the vertex regression errors of $A$ to the $p^\mathrm{th}$ power, minus the vertex regression errors of $B$ to the $p^\mathrm{th}$ power, is $A_p - B_p$.
Since $p <0$ and the $a_i$ and $b_i$ are nondecreasing, this is at least the value one would have if for all $j > i+1$, $b_j = b_{i+1}$ and $a_j$ is extremely large.
In fact, we can let $a_j$ be infinite and $1/a_j = 0$.
Thus $A_p-B_p > a_{i+1}^p - (n-i) b_{i+1}^p$.
As $p\rightarrow 0$, $(a_{i+1}/b_{i+1})^p \rightarrow \infty$ since $a_{i+1} < b_{i+1}$ and $p$ is negative.
Since $(n-i)$ is a constant, for small enough $p$, $a_{i+1}^p > (n-i) b_{i+1}^p$ (recall we were trying to maximize the sum of the errors to the $p^\mathrm{th}$ power).
Thus if $A$ precedes $B$ in the \lex{0}{} ordering it precedes $B$ in the \Strictup{0}{} ordering of regression errors.
This implies that if\lex{0}{} is \Strictup{0}{}.
 
Note that \lex{0}{}, and hence also \Strictup{0}{}, are not unique since for 1, 0 on a linear order both 0, 0 and 1, 1 are optimal.
Nor are they monotonic since their unique regression of 6, 6, 4, 2, 0 is all 6s, while increasing the 2 and 0 values to 4 gives a unique regression of all 4s.

\section{Linear Orders} \label{sec:linear}

\begin{figure}
\centering
\includegraphics[scale=0.7]{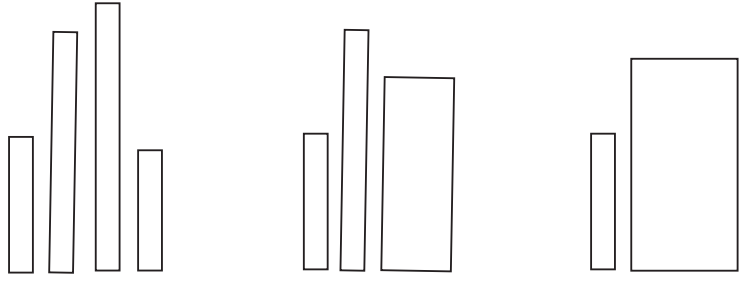}
\caption{Steps in PAV (pool adjacent violators)}    \label{fig:PAVfigure}

\hrulefill
\end{figure}

Our approach for determining \Strictdown{1}{} on a linear order is based on partitioning, using 0-1 isotonic regression to iteratively narrow down to the regression value for each point~\cite{QPartition}.
We describe the process in terms of regression values being in boxes, where the horizontal extent of each box is the interval of vertices it is representing and the vertical extent is guaranteed to include the regression value of every vertex in the box.
For each iteration there are 2 stages: the first partitions each box (or the box may be unchanged), and the second pools (merges) partitions of adjacent boxes to maintain the isotonic property. 
At the end of an iteration the vertical extents of the boxes are consecutive intervals that are the same, followed by another sequence of vertical extents that are higher but all the same, followed by another sequence, etc.
The vertical extents are intervals or single points.
Figure~\ref{fig:PAVfigure} shows a sequence of steps in the pooling stage.

Initially each data point is in its own box, with the vertical extent being the range from smallest to largest data value.
At an intermediate stage, the boxes with vertical extent which is the open interval $(a,b)$ form a sequence (the lower bound $a$ may be included if it is the smallest data value, and similarly for the upper bound $b$).
Let $[i,j]$ be the sequence of index values, i.e., the box is $[i,j] \times (a,b)$.
Denote this box by $B(i,j,a,b)$, and let $W = \sum_{k=i}^j w_k$. 
Let $\hat{y}$ be any value in $(a,b)$, and let
$$
S_\square(i,j,\hat{y}) = \sum \{w_k: i \!\leq\! k \!\leq\! i,\, y(k) \square\, \hat{y}\}~~~~ \mathrm{for~} \square \in \{<, =, >\}
$$
and
$$
P_\square(i,j,\hat{y}) = \prod \{|b-y(k)|^{w_k}: i \!\leq k \!\leq j,\, y(k) \square\, \hat{y}\} ~~~~ \mathrm{for~} \square \in \{ \leq, \geq\}
$$

These values have several properties.
First, for all of the blocks combined they can be computed in $\Theta(n)$ time.
Further, for a given block the $S$ and $P$ values can be used to determine where the median interval is located and its relation to $\hat{y}$.
Also, when two adjacent blocks with the same vertical extent are merged the $S$, $P$, and $W$ values of the resulting block can be computed in constant time from the values of the blocks being merged.
This fact is important because the 2nd step of each stage merges blocks, using \textit{pool adjacent violators} (PAV). 
This is a standard operation in isotonic regression on a linear order, merging adjacent blocks with regression values in the wrong order into a larger block and then determining it's regression value, which will be in the interval of the lower regression value (from the block on the right) to the larger regression value (the block on the left).
In the most common implementation of PAV the blocks are examined in increasing order of the range of their independent variable, and if an out of order pair is discovered they are merged and the interval of regression values determined.
This may be smaller than the preceding block, in which case they are merged, etc.
Fig.~\ref{fig:PAVfigure} illustrates this process, and Fig.~\ref{fig:PAVcode} gives an implementation used for \Strictdown{1}{}.
The total number of mergers throughout the entire algorithm is $ \leq n-1$.

\begin{figure}
\noindent
\Ain{0} If $\hat{y}$ is a data value in $B$ (i.e., $S_= > 0$) then\\
      \Ain{1} if $S_<$ and $S_>$ are $< W/2$ then $\hat{y}$ is the unique median, shrink $B$ to $B^\prime(i,j,\hat{y},\hat{y})$\\
      \Ain{1}else $\hat{y}$ cannot be the desired median\\
           \Ain{2} if $S_< \geq W/2$ then $(a,\hat{y})$ contains the desired median, shrink to $B^\prime(i,j,a,\hat{y})$ \\
           \Ain{2} else $(\hat{y},b)$ contains the median, shrink to $B^\prime(i,j,\hat{y}, b)$\\
\Ain{0} else $\hat{y}$ is not a data value\\
      \Ain{1}if $S_< = S_>$ \{both are $W/2$, $\hat{y}$ is in the median interval, use Jackson's formula\}\\
           \Ain{2} if $P_< = P_>$ shrink to $B^\prime(i,j,\hat{y},\hat{y})$\\
           \Ain{2} else if $P_<  > P_>$ shrink to $B^\prime(i,j,a,\hat{y})$\\
           \Ain{2} else shrink to $B^\prime(i,j,\hat{y},b)$\\
      \Ain{1} else $\hat{y}$ is not in the median interval\\
           \Ain{2} if $S_< > W/2$ shrink to $B^\prime(i,j,a,\hat{y})$ \\
           \Ain{2} else shrink to $B^\prime(i,j,\hat{y},b)$

\caption{Partitioning a block for $\mathsf{strict}{\downarrow}_1$}  \label{fig:partition}

\hrulefill
\end{figure}

\begin{figure}
\noindent
For the sequence of blocks all initially with vertical extent $(a,b)$ in this stage\\
 \\
\Ain{0} Determine $\hat{y}$\\
\Ain{0} Partition the first block \{Figure~\ref{fig:partition}\}\\
\Ain{0} While there are still blocks in this sequence\\
     \Ain{1} Let $B(i,j,a,b)$ be the next block\\
     \Ain{1} Partition $B$, resulting in $B'(i,j,a',b')$~~ \{note that $(a',b')$ is $(a,\hat{y})$, $[\hat{y},\hat{y}]$, or $(\hat{y},b)\}$\\
     \Ain{2} Repeat\\
            \Ain{3} Let $B^*(h,i\!-\!1,a^*,b^*)$ be the predecssor of $B$~ \{it has already been partitioned\}\\
            \Ain{3}  If $(a^*,b^*)$ is above $(a',b')$ then repeat~~ \{the blocks are not isotonic\}\\
               \Ain{4}  reset $B$ to be the block $(h,j,a,b)$ \{i.e., merge the unpartitioned $B$ and $B^*$\},\\
              \Ain{5}  using the $s$, $p$, and $W$ values from $B$ and $B^*$ to calculate those for the new $B$\\
              \Ain{4}  partition $B$, resulting in a new $B'(i,j,a',b')$\\
           \Ain{3} until $(a^*,b^*)$ is not above $(a',b')$\\
 \\
Go to the next sequence of blocks (their vertical extents will be above the current one)

\caption{PAV for a sequence of partitioned blocks with same initial vertical extent}  \label{fig:PAVcode}

\hrulefill
\end{figure}

\begin{table}
\begin{center}

\begin{tabular}{|l|c|c|c|}
\hline        & definition     &  time             & reference\\ \hline
 $L_0$ & \lex{0}{} &  \rule[-0.08in]{0pt}{0.25in}\nlog{3}{0}  & \cite{QL0Secondary}   \\
            & \Strictup{0}{}   &  ?    &   \\ \hline
$L_1$ & \Strictdown{1}{} & \rule[-0.08in]{0pt}{0.25in} $\Theta(n \log (\max \{n, U/\delta\}))$  & Section~\ref{sec:linear} \\ 
           & \Strictup{1}{}   &   ?~~ See note &     \\ \hline
$L_\infty$ & \lex{\infty}{} & \rule[-0.08in]{0pt}{0.25in} \nlog{1}{1}  & \cite{QStrictLinfty} \\
              &  \Strictup{\infty}{} & ?  & \\ \hline
\end{tabular}
\end{center}

\vspace*{-0.15in}
\caption{Fastest Known Isotonic Regression Algorithms  for Weighted Data on a Linear Order}
\centerline{? : Apparently no algorithm is based on this definition}
\centerline{Note: The two definitions do not always produce the same regression}

\hrulefill
\end{table}

An important aspect of Jackson's formula is that the $P$ values are only relevant if $\hat{y}$ is in the interval of median values and the interval has more than one point.
If the interval of medians is $[c,d]$ and $\hat{y} \in [c,d]$ then when $P_\leq(\cdot) = P_\geq(\cdot)$ the optimal median is $\hat{y}$; when $P_\leq(\cdot) > P_\geq(\cdot)$ then the optimal median is $< \hat{y}$; and otherwise is $> \hat{y}$.
During this iteration of the partitioning process $B(i,j,a,b)$ is replaced by $B(i,j,\hat{y},\hat{y})$ (i.e., a vertical extent of a single point)
if $S_<(\cdot)$ and $S_>(\cdot)$ are both $< W/2$ or $\hat{y}$ is one of the median values and $P_\leq(\cdot) = P_\geq(\cdot)$;
else is replaced by $B(i,j,a,\hat{y})$ if $\hat{y}$ is in the interior of the interval of medians and $P_\leq(\cdot) > P_\geq(\cdot)$;
else is replaced by $B(i,j,\hat{y},b)$.

The only part of the algorithm still unspecified is how to choose the $\hat{y}$ values.
There are several options depending on the objectives one wants.
One way is to use data values, initially using a median data value and then at each iteration, for the sequence of blocks with vertical interval $(a,b)$, choosing the data value which is a median of those in that range.
This would take $\Theta(n \log n)$ time, and for each level set of \Strictdown{1}{f} which is a data value the algorithm would identify this as being the level set's value.
Another option is to guarantee that at the end each interval is within $\delta$ of its true value.
In this case one chooses $\hat{y} = (b-a)/2$, continuing this until $b-a \leq \delta$ for every block.
This would take $\Theta(n \log (U/\delta))$ time.
One could combine these, first using partitioning based on data values followed by dividing the range for blocks that do not have a data value as their correct regression value.
In this case the time is $\Theta(n \log (\max \{n, U/\delta\}))$.
One typically assumes that if the largest absolute value of any data value is $U$ then $U/\delta$ grows at most polynomially with $n$, in which case the time is $\Theta(n \log n)$.

\section{Conclusion} \label{sec:Conclude}

For real-valued weighted data on a finite dag $L_p$ isotonic regression is typically defined purely in terms of minimizing the regression error, but here the ``best'' $L_p$ isotonic regressions used additional criteria.
For $1 < p < \infty$ the regression is unique and therefore imposing additional criteria is not useful.
However, for $p \in [0,1] \cup {\infty}$ there may be infinitely many regressions minimizing the regression error, and such criteria helps select among them.

For example, for unweighted data $f$ on a dag $G$ most researchers use the function $g(x) = (\max\{f(y): y \preceq x\} + \min\{f(y): x \preceq y\})/2$ as the optimal $L_\infty$ regression of choice.
While its simplicity and speed of computation recommend it, for data 2, 0, 1 on a linear order this results in 1, 1, 1.5.
In fact, any function of the form 1, 1, $r$ minimizes the $L_\infty$ norm if $r \in [1,2]$.
However, likely many would prefer to use 1, 1, 1.
This is both the regression obtained as the limit, as $p \rightarrow \infty$, of the $L_p$ regression, \Strictup{\infty}{f}, and the unique one defined via a lexical ordering of the regression errors, \lex{\infty}{f} (see Sec.~\ref{sec:definitions}).
The \lex{\infty}{} definition is based on explicitly minimizing large errors, not just minimizing the maximum error, and appears in~\cite{Yale_AllLp,QStrictLinfty}.
The \Strictup{\infty}{} definition has been used repeatedly in various analytical situations, not just isotonic regression, and is known as the Polya approach.
For finite dags the definitions are equivalent but \lex{\infty}{} is more useful in terms of developing efficient algorithms.

There are numerous papers on minimizing $L_1$ error in various settings involving convex cones, but few are as focused as Jackson~\cite{Jackson_Median} in defining the best median value of a set as \wmeandown{1}{}.
The set of isotonic regressions forms a convex cone.
Almost all papers finding $L_1$ regressions on finite sets result in one where all regression values are data values, but for a level set with non-unique median Jackson's value will never be a data value.
Trying to define a best $L_1$ regression as \Strictup{1}{} is not as successful since $L_p$-optimal regressions are not unique when $p<1$, and the objective function is concave, not convex (Sec.~\ref{sec:strictup1}).

Far less attention has been paid to $L_0$, but a natural lexical definition, \lex{0}{}, restricts the regressions minimizing the $L_0$ metric down to a much smaller set.
\lex{0}{} is based on maximizing the number of small errors, not just maximizing the weight of points with 0 error.
A definition of best was also given in terms of limits, \Strictup{0}{}, by extending the minimization of $L_p$ objectives to negative values of $p$.
It was shown that $\Strictup{0}{} = \lex{0}{}$ (Sec.~\ref{sec:L0}).

Many others have studied related problems such as the rate of convergence of \Strictup{\infty}{} and \Strictdown{1}{}~\cite{EggerTaylorPolya1ConvRate93,QFOMMB_PolyaL1}, other ways to select subsets of $L_0$ isotonic regressions with desired properties~\cite{RadeBM_Relabel09,QL0Secondary}, generating all $L_0$ isotonic regressions~\cite{StegemanFeeldersAllL0_11} or their core~\cite{RadeBM_Relabel12}, and finding a minimal $L_1$ isotonic regression~\cite{ShiMinL195}.

There are several open questions concerning algorithms for the problems studied here.
For example, it is known how to use maximal flow algorithms to find $L_0$ isotonic regressions of weighted data on arbitrary dags
\cite{Relab_Feelders06,PP_Relabel,RadeBM_Relabel09,QL0Secondary}.
However, these algorithms are only guaranteed to produce $L_0$ isotonic regressions, not \lex{0}{} regressions.
It would be interesting to find a more efficient algorithm for general dags.
For $L_1$, for general dags the algorithms for finding isotonic regressions do not always produce \Strictdown{1}{}, and it would be useful to find an efficient algorithm that produces \Strictdown{1}{} directly, rather than through approximation as in Sec.~\ref{sec:strictdown1}.

\section*{Appendix}

In general there are widely varying $L_\infty$ isotonic regressions of a specific dag and data, and in~\cite{QStrictLinfty} there are characterizations of various properties of $L_\infty$ isotonic regression algorithms.
Two properties of particular interest are monotonicity and maintaining level set trimming.
For an $L_\infty$ isotonic regression algorithm $\mathcal{A}$, for dag $G=(V,E)$ and weighted function $(f,w)$ on $G$
let $\mathcal{A}(G,f,w)$ denote the isotonic regression that $\mathcal{A}$ produces.
$\mathcal{A}$ is an \textit{$L_\infty$ isotonic operator on $G$} if $\mathcal{A}(G,f,w)$ is an $L_\infty$ isotonic regression of $(f,w)$ for all $(f,w)$.

$\mathcal{A}$ is \textit{monotonic} on $G$ iff for all weight functions $w$ and functions $f$ and $g$ on $V$, if $f$ is pointwise less than or equal to $g$ on $V$ then $\mathcal{A}(G,f,w)$ is pointwise less than or equal to $\mathcal{A}(G,g,w)$.
Algorithm $\mathcal{A}$ \textit{preserves level set trimming} on $G$ iff for any weighted function $(f,w)$ on $G$, for any level set \textsf{L} of $\mathcal{A}(G,f,w)$, the regression values on \textsf{L} are \wmeanup{\infty}{(f,w)|L}

In~\cite{QStrictLinfty} it was shown that \Strictup{\infty}{} is monotonic and preserves level set trimming for all dags.
The following shows that this characterizes \Strictup{\infty}{}, in that any $L_\infty$ isotonic regression algorithm which is monotonic and preserves level set trimming of weighted data functions on a dag $G$ always produces \Strictup{\infty}{} on $G$.
It does not appear in~\cite{QStrictLinfty} since the author only noticed it
after that paper was in the publication processes.

\begin{theorem}
For any DAG $G = (V,E)$ and $L_\infty$ isotonic regression algorithm $\mathcal{A}$ on $G$, if $\mathcal{A}$ is monotonic and
preserves level set trimming then $\mathcal{A}$ always produces \Strictup{\infty}{} (and hence \lex{\infty}{}) on $G$.
\end{theorem}
\textbf{Proof:}
We use proof by contradiction.
Suppose $\mathcal{A}$ is monotonic and preserves level set trimming, and there
is a weighted function $(f,w)$ for which $\mathcal{A}(G,f,w) \neq \Strictup{\infty}{G,f,w}$.
Among the level sets of \Strictup{\infty}{G,f,w} which are not level sets of $\mathcal{A}$, or where the
regression values differ, let $L$ be one of maximal error.
Let $f_1$ be $f$ trimmed on all level sets of \Strictup{\infty}{G,f,w}
with error greater than $L$'s.
Then $\Strictup{\infty}{G,f_1,w} \!=\! \Strictup{\infty}{G,f,w}$,\, $\mathcal{A}(G,f_1,w)\!=\!\mathcal{A}(G,f,w)$,
and the $L_\infty$ regression error of $\Strictup{\infty}{G,f_1,w}$ is its $L_\infty$ error on $L$ (there may be other
level sets with the same error).
Let $c$ be the value of \Strictup{\infty}{G,f_1,w} on $L$.
If $\mathcal{A}(G,f_1,w)$ does not equal $c$ on all of $L$ then it has larger regression error
than $\Strictup{\infty}{G,f_1,w}$, in
which case it is not an isotonic regression.
Otherwise, it has a level set
$B \varsupsetneq L$ with regression value $c$.
Let $B^\prime = \{u: u \in B,\,
  \Strictup{\infty}{G,f_1,w}(u) < c\}$
(if $B^\prime$ is empty then a similar proof can be applied to the set where
$\Strictup{\infty} > c$).
Since $\Strictup{\infty}{(f_1,w)} < c$ on $B^\prime$, and raising the values to $c$ would not violate
the isotonic condition, it must be that $\setmean{B^\prime} < c$.

Let $f_2$ be the function formed by trimming $f_1$ on all level sets of $\mathcal{A}(G,f_1,w)$ except $B$.  
Then $\mathcal{A}(G,f_2,w) = \mathcal{A}(G,f_1,w)$.
Define $f_3(u)$ to be $f_2(u)$ if $f_2(u) < c$ or 
$u \in B^\prime$,
and $M$ otherwise, where $M$ is the maximum value of $f_2$.
Since $\mathcal{A}(G,f_2,w) = c$ on $B^\prime$ and pointwise $f_3 \geq f_2$, 
by monotonicity 
$\mathcal{A}(G,f_3,w) \geq c$ on $B^\prime$.
Let $h$ be the function where $h(u) = f_3(u)$ on $V \setminus B^\prime$, and
$h(u) = \max\{\setmean{B^\prime},D\}$ on $B^\prime$, where $D$ is the largest value of
$f_3$ less than $c$.
Then $h$ is isotonic, and as a regression of $f_3$
has no error on $V \setminus B^\prime$
and smaller $L_\infty$ error than $\mathcal{A}(G,f_3,w)$ on $B^\prime$.
Thus $\mathcal{A}(G,f_3,w)$ is not optimal and hence $\mathcal{A}$ is not an $L_\infty$
regression operator on $G$.\\
$\Box$

\end{document}